\documentclass[conference]{IEEEtran}
\IEEEoverridecommandlockouts
\usepackage{cite}
\usepackage{amsmath,amssymb,amsfonts,dsfont}
\usepackage{algorithmic}
\usepackage{graphicx}
\usepackage{textcomp}
\usepackage{xcolor}
\usepackage{booktabs}
\usepackage{caption}
\usepackage{subcaption}

\usepackage[all=normal,paragraphs=tight,floats=normal,mathspacing=normal,wordspacing=tight,charwidths=tight,mathdisplays=normal,leading=tight]{savetrees}

\newcommand{\abs}[1]{\ensuremath{\left| #1 \right|}}

\DeclareMathOperator*{\argmax}{arg\,max}

\newcommand{\norm}[2]{\ensuremath{\left\lVert #1 \right\rVert}_#2}

\newcommand*{\herm}{{\mathsf{H}}}

\def\BibTeX{{\rm B\kern-.05em{\sc i\kern-.025em b}\kern-.08em
    T\kern-.1667em\lower.7ex\hbox{E}\kern-.125emX}}

\begin{document}

\title{Model-based Deep Learning for Beam Prediction based on a Channel Chart}

\author{
	\IEEEauthorblockN{
        Taha Yassine\IEEEauthorrefmark{1}$^,$\IEEEauthorrefmark{2}$^,$$^\star$,
        Baptiste Chatelier\IEEEauthorrefmark{3}$^,$\IEEEauthorrefmark{2}$^,$\IEEEauthorrefmark{1}$^,$$^\star$,
        Vincent Corlay\IEEEauthorrefmark{3}$^,$\IEEEauthorrefmark{1},
        Matthieu Crussière\IEEEauthorrefmark{2}$^,$\IEEEauthorrefmark{1},
        Stéphane Paquelet\IEEEauthorrefmark{1},\\
        Olav Tirkkonen\IEEEauthorrefmark{4},
		Luc Le Magoarou\IEEEauthorrefmark{2}$^,$\IEEEauthorrefmark{1}
		}
	\IEEEauthorblockA{
		\IEEEauthorrefmark{1}b\raisebox{0.2mm}{\scalebox{0.7}{\textbf{$<>$}}}com, Rennes, France
		}
	\IEEEauthorblockA{
		\IEEEauthorrefmark{2}Univ Rennes, INSA Rennes, CNRS, IETR-UMR 6164, Rennes, France
	}
	\IEEEauthorblockA{
		\IEEEauthorrefmark{3} Mitsubishi Electric R\&D Centre Europe, Rennes, France
	}
	\IEEEauthorblockA{
		\IEEEauthorrefmark{4} Department of Information and Communications Engineering, Aalto University, Finland}
    $^\star$ \textbf{\textit{Equal contribution}}
	}
\maketitle

\begin{abstract}
    Channel charting builds a map of the radio environment in an unsupervised way. The obtained chart locations can be seen as low-dimensional compressed versions of channel state information that can be used for a wide variety of applications, including beam prediction. In non-standalone or cell-free systems, chart locations computed at a given base station can be transmitted to several other base stations (possibly operating at different frequency bands) for them to predict which beams to use. This potentially yields a dramatic reduction of the overhead due to channel estimation or beam management, since only the base station performing charting requires channel state information, the others directly predicting the beam from the chart location. In this paper, advanced model-based neural network architectures are proposed for both channel charting and beam prediction. The proposed methods are assessed on realistic synthetic channels, yielding promising results.
\end{abstract}


\IEEEpeerreviewmaketitle

\section{Introduction}
\label{sec:introduction}

\IEEEPARstart{L}{arge} amounts of available bandwidth in the millimeter wave (mmWave) frequency bands is a key enabler of the high throughput requirements of Fifth Generation (5G) and Beyond 5G (B5G) communication systems. Such high frequencies allow antenna sizes to decrease, resulting in practical massive Multiple Input Multiple Output (mMIMO) systems. As such systems possess high spatial directivity, it is of paramount importance to have precise beam management methods to benefit from the associated beamforming gain. Indeed, in mMIMO systems, a small angular error in the base station's (BS) beam steering angle towards the User Equipment (UE) could lead to a highly degraded link performance~\cite{Chatelier2022}. In order to keep beam selection tractable, 3GPP standardized beam management procedures for 5G New Radio (5GNR)~\cite{3gpp.38.214}. The standard imposes for each BS to possess a codebook of precoders from which it picks the one that maximizes link performance for each individual UE. This procedure has to be carried out frequently as the UEs are likely to be moving. All of those constraints result in a complex beam management procedure, requiring constant exchanges between all BSs and UEs.

In the last decade, machine learning (ML) has emerged as a promising solution in many communication problems such as channel estimation~\cite{Hengtao18,Balevi2020,yassine2022,Chatelier2023,chatelier2023modelbased}, detection~\cite{Samuel2019,Corlay2018}, positioning~\cite{Chatelier2023b} or decoding~\cite{Corlay2022}. ML methods have been successfully applied for location-based beamforming~\cite{LeMagoarou2022} and beam management~\cite{Ponnada2021,Kazemi2023}. More particularly, approaches combining channel charting and deep learning have been used for beam prediction and precoder learning in non-standalone mmWave~\cite{Ponnada2021} and Cell Free (CF) mMIMO systems~\cite{LeMagoarou2022a}. Channel charting is a dimensionality reduction method for channel vectors: it yields a pseudo-location (i.e. compressed channel) map from channels in an unsupervised manner.

In CF mMIMO systems, a UE can be served by multiple BSs. In such scenarios, one can think of the following approach: a BS generates a pseudo-location from a collected uplink channel and sends this pseudo-location to other BSs. Then, at each BS, a neural network is used to infer the best beam in its codebook, based solely on the received pseudo-location. This approach only requires to estimate the uplink channel at one BS, resulting in a great reduction of beam management complexity. 

\noindent\textbf{Contributions.} In this paper, a model-based neural network is introduced to learn the mapping from a pseudo-location to the best beam in a codebook. This can be seen as a classification network. The proposed approach allows to greatly reduce the beam management complexity compared to classical methods. Classification based on discrete codebooks is also compared to a regression approach, where a precoder is directly learned without constraining it to come from a given codebook. Performance is assessed on realistic synthetic channels obtained via the Sionna~\cite{Sionna} and DeepMIMO~\cite{DeepMIMO} datasets.

\noindent\textbf{Related work.} Inferring a beam in a codebook from a pseudo-location has recently been proposed in~\cite{Ponnada2021}. However, this paper used a classical multilayer perceptron (MLP) for the beam prediction network and only considered a small scene for the UE locations. Directly learning a precoder has also been studied in~\cite{Maiberger2010,LeMagoarou2022,LeMagoarou2022a}. More particularly,~\cite{LeMagoarou2022a} proposed a model-based neural network for the precoder learning task. However in~\cite{LeMagoarou2022a}, directly learned precoders were not compared to a codebook-based classification approach.

\section{Problem formulation}
\label{sec:problem}
%Let us define the system model and clearly express the objective of this study.

\begin{figure}[tb]
	\centering
	\includegraphics[scale=.45]{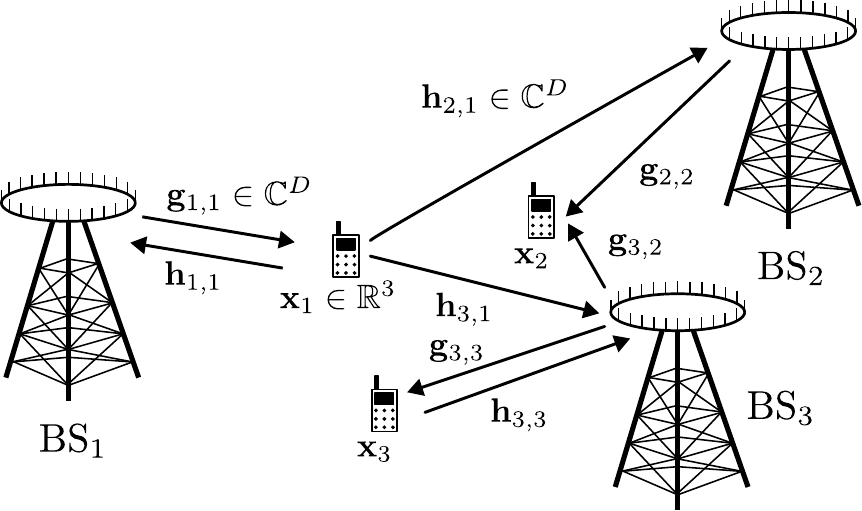}
	\caption{System model ($B=3, N_u = 3$)}
	\label{fig:system_model}
\end{figure}

%\noindent\textbf{System model.} 
In this paper, a CF mMIMO system is considered, as depicted on Fig.~\ref{fig:system_model}, comprising $B$ BSs using uniform planar arrays (UPAs) with $N_a$ antennas, and $N_u$ single-antenna UEs. Multicarrier transmissions with $N_s$ subcarriers are considered. Note that the approach could be easily transposed with multi-antenna UEs as in~\cite{Ponnada2021}. The system operates with different uplink and downlink central frequencies $f_{\text{ul}}$ and $f_{\text{dl}}$. Let $\mathbf{h}_{i,j}\in \mathbb{C}^D$ (resp. $\mathbf{g}_{i,j}\in \mathbb{C}^D$) be the uplink (resp. downlink) channel between BS $i$ and UE $j$. Note that vectorized channels are considered, hence $D=N_a N_s$. The location of UE $j$ is denoted as $\mathbf{x}_j \in \mathbb{R}^3$.

% Presentation codebook 
Each BS uses the same $2$D Discrete Fourier Transform (DFT) codebook $\mathbf{C} \in\mathbb{C}^{N_b\times N_a}$. It is defined, using the Kronecker product, as $\mathbf{C} = \frac{1}{\sqrt{N_v N_h}}\mathbf{\Psi}_v \otimes \mathbf{\Psi}_h$. $\mathbf{\Psi}_v \in \mathbb{C}^{O_v N_v \times N_v}$ and $\mathbf{\Psi}_h \in \mathbb{C}^{O_h N_h \times N_h}$, where $N_h$ and $O_h$ (resp. $N_v$ and $O_v$) represent the number of antennas and oversampling factor for the horizontal (resp. vertical) dimension of the BS UPA. Note that the number of beams in the 2D-DFT codebook is a multiple of the number of BS antennas: $N_b = O_v O_h N_v N_h = O_v O_h N_a$, as $N_a = N_v N_h$. In this paper, $O_h = O_v = 2$ so that the oversampling factor is uniformly split between the azimuth and elevation dimensions. For $k=h,v$, the 1D-DFT codebook can be defined as follows:
\begin{equation}
	\mathbf{\Psi}_k = \begin{bmatrix}
		1 & 1 & \cdots & 1\\
		1 & \mathrm{e}^{\frac{\mathrm{j}2\pi}{O_k N_k}} & \cdots & \mathrm{e}^{\frac{\mathrm{j}2\pi\left(N_k - 1\right) }{O_k N_k} }\\
		\vdots & & & \\
		1 & \mathrm{e}^{\frac{\mathrm{j}2\pi\left(O_k N_k - 1\right)}{O_k N_k}} & \cdots & \mathrm{e}^{\frac{\mathrm{j}2\pi\left(O_k N_k - 1\right)\left(N_k - 1\right) }{O_k N_k} }
	\end{bmatrix}.
\end{equation}

For each UE $u$, the best beam in the codebook at BS $b$ is:
\begin{equation}\label{eq:eq_2}
	\mathbf{c}_{i^\star_{b,u}} \in \mathbb{C}^{N_a} \text{ with } i^\star_{b,u} = \argmax_{i} \dfrac{1}{N_s} \sum_{s=1}^{N_s} \dfrac{\abs{\mathbf{c}_i^{\herm}\tilde{\mathbf{g}}_{b,u,s}}^2}{\norm{\tilde{\mathbf{g}}_{b,u,s}}{2}^2},
\end{equation}
where $\tilde{\mathbf{g}}_{b,u,s} \in \mathbb{C}^{N_a}$ is the downlink channel between BS $b$ and UE $u$ at subcarrier $s$. In other words, the chosen precoder maximizes the mean correlation over all subcarriers between the precoder and the downlink channel. The $k$th best beam index at BS $b$ for UE $u$ is denoted by $i^{\star,k}_{b,u}$.

\noindent\textbf{Objective.} The pursued objective is to choose an appropriate precoder based on a low-dimensional representation of the channel, possibly computed at another BS and/or at a different frequency. To do so, channel charting is used to obtain a pseudo-location $\mathbf{z} \in \mathbb{R}^d$ from an uplink channel $\mathbf{h} \in \mathbb{R}^D$ estimated at one BS. Then, that pseudo-location is sent to all the other BSs that can then infer the best beam in each BS codebook from that pseudo-location. Formally, the proposed contribution can be defined as an encoding function whose role is to compress channels (via channel charting):
\begin{equation}
	\begin{aligned}
		\mathcal{C}\colon \mathbb{C}^D &\longrightarrow \mathbb{R}^d\\
		\mathbf{h} &\longrightarrow \mathbf{z} \triangleq \mathcal{C}\left(\mathbf{h}\right),
	\end{aligned}
\end{equation}
and a decoding function whose role is to map compressed channels to appropriate beams (beam selection), which solves a classification problem:
\begin{equation}
	\begin{aligned}
		\mathcal{D}\colon \mathbb{R}^d &\longrightarrow \mathbb{N}^*\\
		\mathbf{z} &\longrightarrow i \triangleq \mathcal{D}\left(\mathbf{z}\right).
        \label{eq:decoder}
	\end{aligned}
\end{equation}
Our intention is to compare the proposed technique with the one presented in~\cite{LeMagoarou2022a}, where a precoder (not constrained to belong to a codebook) is inferred from the pseudo-location. In that case the decoding function becomes a precoder decoding function, which solves a regression problem:
\begin{equation}
	\begin{aligned}
		\mathcal{P}\colon \mathbb{R}^d &\longrightarrow \mathbb{C}^{N_a}\\
		\mathbf{z} &\longrightarrow \mathbf{w}\triangleq  \mathcal{P}\left(\mathbf{z}\right).
	\end{aligned}
\end{equation}

Note that the precoder is learned only for one subcarrier (e.g. the central subcarrier): it is possible to learn it for every subcarrier but it is not done in this paper for the sake of clarity in the comparisons. One can remark that, due to Eq.~\eqref{eq:eq_2}, the inferred beam by the classification learner is valid for all subcarriers. This also holds true for the regression learner under the narrowband assumption. Note that there is only one encoding function at the BS doing the channel charting, but $B$ decoding functions at every BS.

\noindent\textbf{Performance measures.} In order to measure the performance of the beam decoding function, i.e. decoding the correct beam index from the pseudo-location at a particular BS $b$, it is proposed to use the modified top-$k$ accuracy defined as:
\begin{equation}
	\gamma_b^k = \dfrac{1}{N_u}\sum_{u=1}^{N_u} \mathds{1}_{\left\{\mathcal{D}\left(\mathbf{z}_u\right)_b \in \left\{i^{\star}_{b,u},i^{\star,2}_{b,u},\cdots,i^{\star,k}_{b,u}\right\}\right\}}.
\end{equation}
This metric is between $0$ and $1$: $0$ means that predicted beams are never among the $k$ best, while $1$ means that all predicted beams are always among the $k$ best. When learning the precoder, the performance measure at BS $b$ for each UE $u$ is the normalized correlation between the learned precoder and the downlink channel (at the central subcarrier denoted $f$), as proposed in~\cite{LeMagoarou2022}:
\begin{equation}\label{eq:corr}
	\eta_{b,u} =\dfrac{\abs{\mathcal{P}\left(\mathbf{z}_u\right)_b^{\herm}\tilde{\mathbf{g}}_{b,u,f}}^2}{\norm{\tilde{\mathbf{g}}_{b,u,f}}{2}^2}.
\end{equation}
This metric is between $0$ and $1$: $0$ means that the precoder is orthogonal to the downlink channel, while $1$ means that it is perfectly aligned (collinear) with the downlink channel.
\section{Proposed method}
\label{sec:method}
This section covers the proposed beam selection method that can be viewed in two separate stages: inference and training. A similar approach as in~\cite{LeMagoarou2022a,yassine_spawc} is used in this paper. As exposed earlier, the encoding stage is done using channel charting only at one BS, called BS$1$ for the rest of the paper.

\begin{figure}[!h]
	\centering
	\includegraphics[scale=.4]{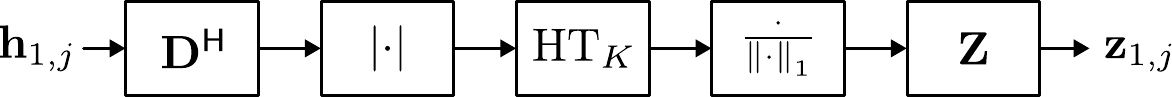}
	\caption{Method for out-of-sample $\mathbf{h}_{1,j}$ channel charting~\cite{LeMagoarou2022a,yassine_spawc}}
	\label{fig:inference_isomap}
\end{figure}

\noindent\textbf{Inference.} The inference phase of the proposed method can be split into three distinct steps:
\begin{itemize}
	\item Using the \texttt{ISOMAP} algorithm as the channel charting method, $d$ latent variables ($d \ll D$) are computed from an uplink channel $\mathbf{h}_{1,j} \in \mathbb{C}^D$ to form the pseudo-location $\mathbf{z}_{1,j} \in \mathbb{C}^d$ at BS$1$. As it is conventional for manifold learning methods, \texttt{ISOMAP} is computationally intensive in out-of-sample scenarios. As this inference method is used for the beam management of each new UE, it is of paramount importance to optimize it. A way to overcome this complexity issue has been presented in~\cite{LeMagoarou2022a} and is recalled in Fig.~\ref{fig:inference_isomap}. The new pseudo-location $\mathbf{z}_{1,j}$ can be seen as a wisely chosen convex combination of the pseudo-locations obtained with the calibration uplink channels.
	\item The computed pseudo-location  $\mathbf{z}_{1,j}$ is sent to all other BSs by BS$1$.
	\item All BSs use the received pseudo-location to perform beam selection using their decoding functions.
\end{itemize}

A beam management complexity comparison is presented in Table~\ref{table:inf_complexities}. The classical approach is beam sweeping: each BS has to test each beam of its codebook, meaning that for each of the $B$ BSs, the estimation of $D$ complex numbers is required. On the other hand, the proposed approach only requires one uplink channel estimation, and the transmission of the computed pseudo-location, of dimension $d$, to the other BSs, achieving a substantial complexity reduction.

\begin{table}[!h]
	\centering
	\begin{tabular}{lcc}
		\toprule
		& Beam sweeping & Proposed method \\
		\midrule
		Inference complexity & $\mathcal{O}\left(B D\right)$ & $\mathcal{O}\left(D+Bd\right)$\\
		\bottomrule
	\end{tabular}
	\caption{Inference complexities ($d\ll D$)}
	\label{table:inf_complexities}
\end{table}

\noindent\textbf{Training.}
The main objective of this contribution is to propose efficient encoding and decoding functions. For the encoding function, as in~\cite{LeMagoarou2022a}, the dimensionality reduction algorithm \texttt{ISOMAP} is used to achieve channel charting. An uplink channel calibration dataset $\left\{\mathbf{h}_{1,n}\right\}_{n=1}^{N_{\text{cal}}}$ is collected at BS$1$. Then, the \texttt{ISOMAP} algorithm is applied using the phase-insensitive distance proposed in~\cite{LeMagoarou2021}. This results in a pseudo-location calibration dataset $\left\{\mathbf{z}_{1,n}\right\}_{n=1}^{N_{\text{cal}}}$. Two matrices are then defined: $\mathbf{D} = \left(\mathbf{h}_{1,1},\cdots,\mathbf{h}_{1,N_{\text{cal}}}\right) \in \mathbb{C}^{D\times N_{\text{cal}}}$, and $\mathbf{Z} = \left(\mathbf{z}_{1,1},\cdots,\mathbf{z}_{1,N_{\text{cal}}}\right) \in \mathbb{C}^{d\times N_{\text{cal}}}$. Those matrices are used in the fast out-of-sample channel charting procedure presented in Fig.~\ref{fig:inference_isomap}.
It is proposed to use a neural network to learn the encoding function: the input is a pseudo-location and the output is a probability vector representing the probability of selection for each beam in the codebook. Note that, in contrast to Eq.~\eqref{eq:decoder}, the output of the decoder is not a scalar. This is simply an implementation detail, as computing the argmax of this probability vector gives the best beam index.

\begin{figure}[!h]
	\centering
	\includegraphics[scale=.9]{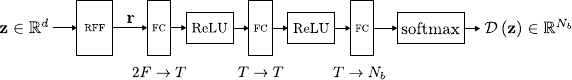}
	\caption{Proposed RFF architecture}
	\label{fig:RFF_net}
\end{figure}

As the spatial best beam distribution function contains high frequencies, i.e. there exist regions (e.g. close to the BS) for which the best beam varies rapidly with the location, one can use a random Fourier features (RFF) network architecture~\cite{Rahimi2007,Tancik2020}. Indeed, it has been shown that classical MLP networks are biased towards learning low frequency functions~\cite{Rahaman2019,Cao2021}. In order to overcome that issue, new architectures such as the RFF have been proposed. The proposed network architecture can be seen in Fig.~\ref{fig:RFF_net}, where:
\begin{equation}
	\mathbf{r} = \begin{bmatrix}
		\cos\left(2\pi \mathbf{B}\mathbf{z}\right)\\
		\sin\left(2\pi \mathbf{B}\mathbf{z}\right)
	\end{bmatrix},
\end{equation}
with $\mathbf{B}\in \mathbb{R}^{F\times d}$ initialized as $\mathbf{B} \sim \mathcal{N}\left(\mathbf{0}_{F},\sigma^2 \mathbf{Id}_F\right)$ where $\sigma$ is a hyperparameter controlling the frequency range of the RFF embedding layer: the higher the $\sigma$ the easier it will be for the network to learn high frequency content. Note that $\mathbf{B}$ is learned during training.

During training, the loss function to minimize at BS $b$, with batch size $\mathcal{B}_s$, is the multiclass cross-entropy defined as:
% \begin{equation}
% 	\mathcal{L} = -\sum_{u=1}^{\mathcal{B}_s} \mathbf{p}_{b,u} \log_2 \hat{\mathbf{p}}_{b,u}
% \end{equation}
\begin{equation}
	\mathcal{L}_b = -\sum_{u=1}^{\mathcal{B}_s} \mathbf{p}_{b,u} \log_2 \mathcal{D}\left(\mathbf{z}_u\right),
\end{equation}
where $\mathbf{p}_{b,u} \in \left[0,1\right]^{N_b}$ is the true beam probability vector for UE $u$ at BS $b$, i.e. $\left(\mathbf{p}_{b,u}\right)_l = 1 \Leftrightarrow i^\star_{b,u} = l$.

\section{Experiments}
\label{sec:expe}
\noindent\textbf{Simulation settings.} The experiments are carried out on the same dataset as in~\cite{LeMagoarou2022a}. Namely, the DeepMIMO~\cite{DeepMIMO} urban outdoor `O$1$' dataset is used, with $B=2$ BSs equipped with UPAs with $N_a = 64$ antennas. $N_s = 16$ subcarriers are considered over a $20$ MHz bandwidth in both the uplink and downlink.
Another dataset is generated with the same parameters, using the Sionna~\cite{Sionna} ray-tracing library. This dataset is generated in the \textit{Etoile} scene in Paris, France. For both scenarios, the uplink central frequency $f_{\text{ul}}$ is $3.5$ GHz, while the downlink central frequency $f_{\text{dl}}$ is $28$ GHz. In the DeepMIMO scene, there are $N_u = 4.2$k training UEs (UE spatial density: $0.17$ UE/$m^2$), while there are $N_u = 7$k training UEs in the Sionna scene (UE spatial density: $0.042$ UE/$m^2$). One can remark that the training UE spatial density is quite low, making it similar to what could be encountered in actual systems.

\noindent\textbf{Network parameters and baselines.} For the proposed architecture, $F=200$ frequencies are considered, and the hidden layer size is $T=64$. The baseline for comparison is an MLP neural network (with the same number of parameters as the proposed architecture): it is the same network as in Fig.~\ref{fig:RFF_net}, but the input layer is replaced by a classical fully connected layer. Another baseline is the Nearest-Neighbour ($1$-NN) ML method: for a given test pseudo-location, the $1$-NN predicted beam will be the best beam of the nearest train pseudo-location.

\begin{figure*}[!t]
\centering
\begin{minipage}{.5\textwidth}
    \centering
    \includegraphics[scale=.4]{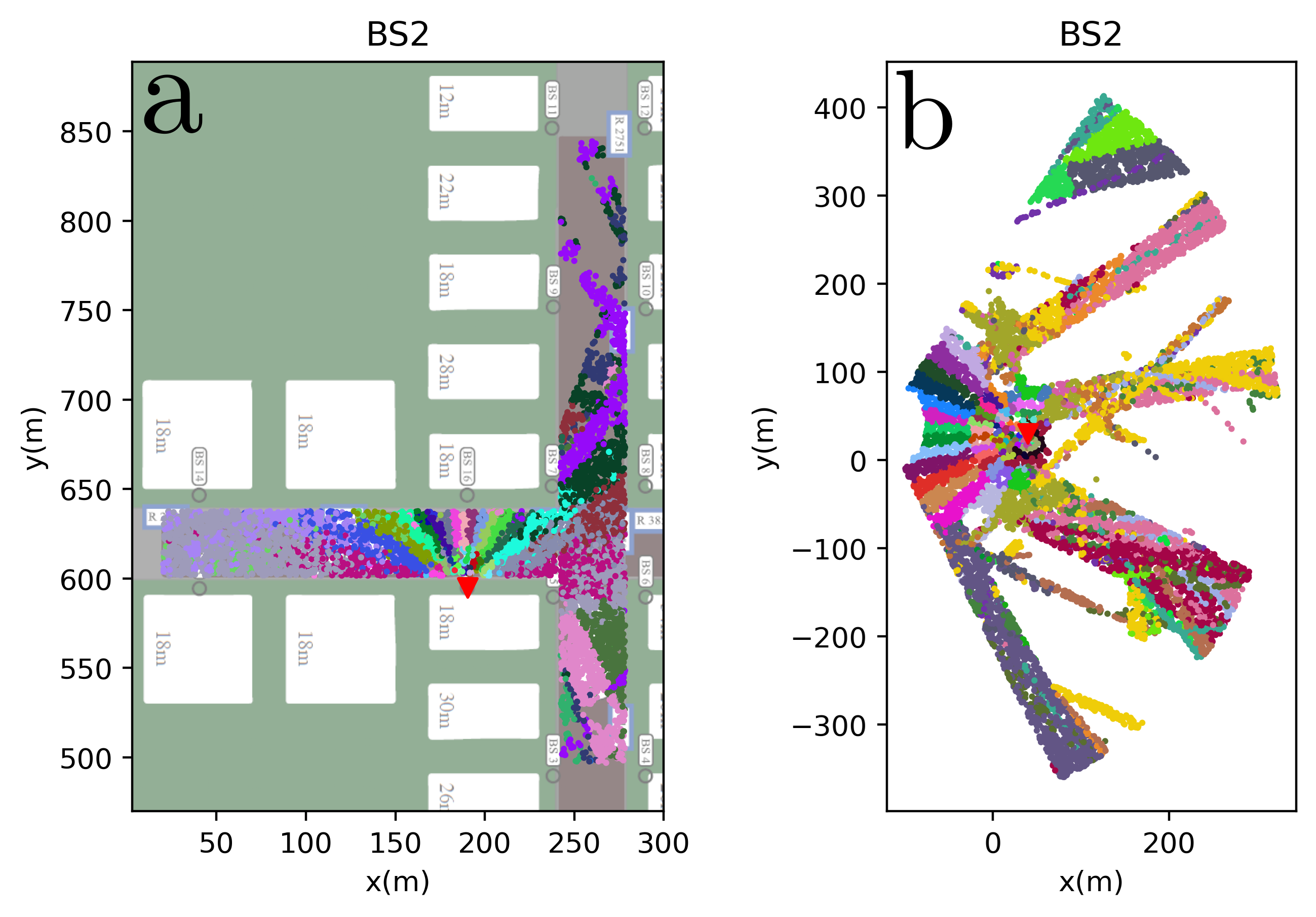}
    \caption{Best beam spatial distribution (a: \textit{DeepMIMO}, b: \textit{Sionna})}
    \label{fig:best_beam_spatial_rep}
\end{minipage}%
\begin{minipage}{.5\textwidth}
	\centering
	\includegraphics[scale=.4]{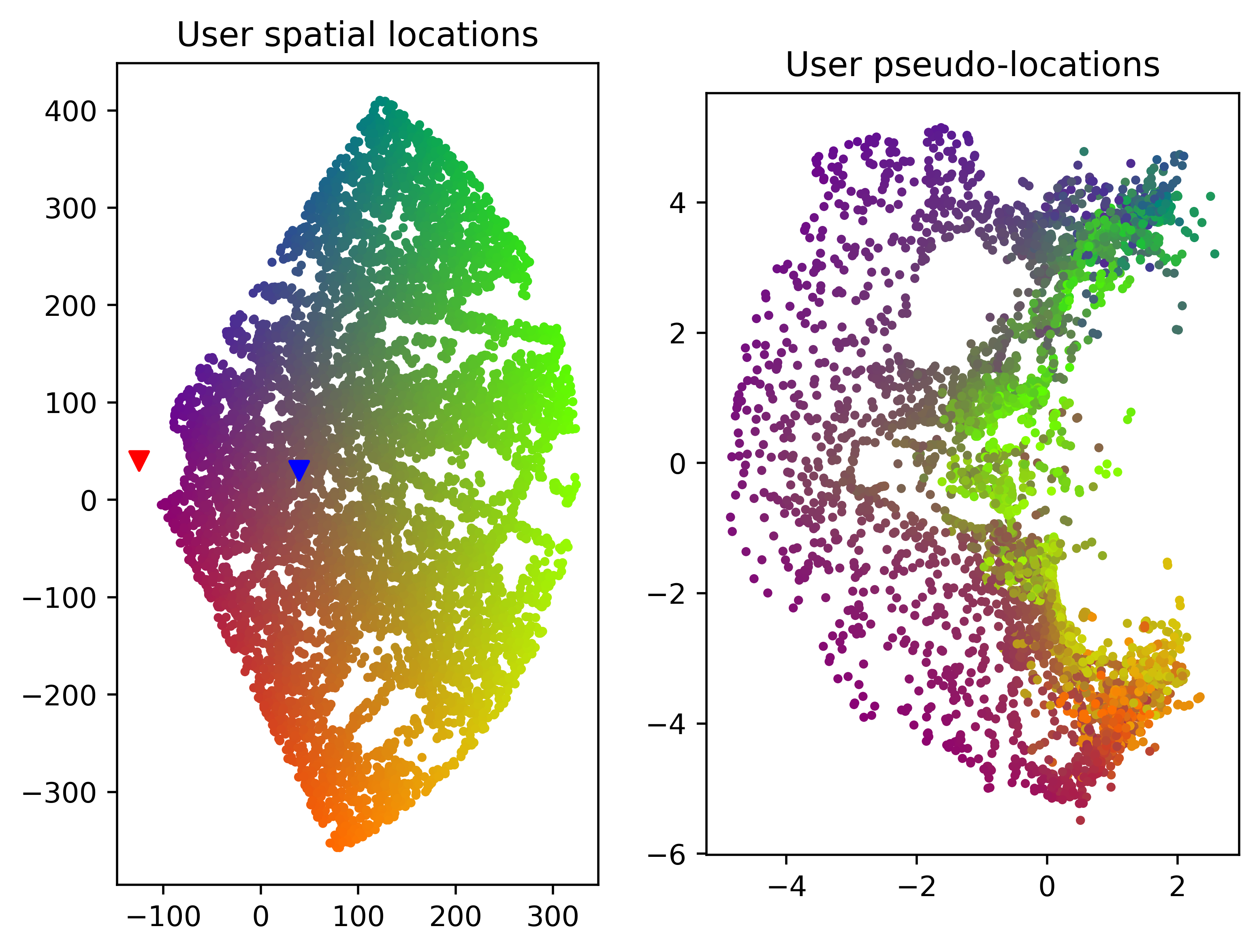}
	\caption{Channel charting map (\textit{Sionna})}
	\label{fig:charting_sionna}
\end{minipage}
\end{figure*}

\noindent\textbf{Best beam spatial distribution.} In Fig.~\ref{fig:best_beam_spatial_rep}, the best beam spatial distribution for the DeepMIMO and Sionna datasets is depicted. Each precoder in the codebook ($N_b=4N_a=256$) is color-coded with a unique color. One can see that higher spatial frequencies appear in the DeepMIMO dataset: near BS$2$ (red triangle), the best beam changes very quickly.

% \begin{figure}
% 	\centering
% 	\includegraphics[scale=.4]{figs/joint_beam.png}
% 	\caption{Best beam spatial distribution (a: \textit{DeepMIMO}, b: \textit{Sionna})}
% 	\label{fig:best_beam_spatial_rep}
% \end{figure}

\noindent\textbf{Charting performance.} The channel charting map can be seen in Fig.~\ref{fig:charting_sionna}. One can remark that UEs close to BS$1$ (UEs in purple), appear more spaced-out in the pseudo-location space: this phenomenon is due to the fact that the angular resolution of antenna arrays is independent of the distance, but the tangential displacement corresponding to a given angular deviation is proportional to the distance (ways to quantify this precisely are detailed in~\cite{yassine2023optimizing}). Channel charting metrics for both datasets are presented in Table~\ref{table:charting_metrics} (see~\cite{LeMagoarou2021,Kazemi2023} for proper definitions). All metrics: Trustworthiness (TW), Continuity (CT) and Kruskal-Stress (KS) are within the $\left[0,1\right]$ range. Both TW and CT are optimal at $1$ while KS is optimal at $0$. Globally, those metrics quantify how well the global structure and neighborhoods in the location space are preserved in the pseudo-location space. One can remark in Table~\ref{table:charting_metrics} that, for both datasets, TW and CT are very good, while KS is good for the Sionna dataset and average for the DeepMIMO dataset.

% \begin{figure}[!h]
% 	\centering
% 	\includegraphics[scale=.4]{figs/sionna_chart.png}
% 	\caption{Channel charting map (\textit{Sionna})}
% 	\label{fig:charting_sionna}
% \end{figure}

\begin{table}[!h]
	\centering
	\begin{tabular}{lccc}
		\toprule
		& TW & CT & KS\\
		\midrule
		\textit{DeepMIMO} & $0.973$ & $0.929$ & $0.471$\\
		\midrule
		\textit{Sionna} & $0.960$ & $0.952$ & $0.292$\\
		\bottomrule
	\end{tabular}
	\caption{Charting metrics (5\% neighbourhoods)}
	\label{table:charting_metrics}
\end{table}

\noindent\textbf{Beam prediction performance.} The beam prediction performance at BS$2$ (the one not used for channel charting) can be found in Table~\ref{table:res_acc_pseudolocs}. Note that the proposed RFF network always outperforms the classical MLP. It is worth noting that the $1$-NN approach works well in the Sionna dataset, with slightly better performances than the RFF network. It is interesting as it shows that the information contained in the low dimensional pseudo-location is sufficient for very a simple ML method such as the 1-NN to have good performance. One can remark that, even if the top-$1$ accuracy performance is only passable, the top-$3$ performance is good in both datasets. This means that using the pseudo-locations allows to have good performance in predicting one of the three best beams in the codebook. This does not result in a huge performance loss due to the high beam density in a $256$ beams $2$D-DFT codebook: the second and third best beams still have high correlations.

\begin{table}[!h]
	\centering
		\begin{tabular}{lccc}
			\toprule
			\textit{DeepMIMO} &  RFF & MLP & $1$-NN\\
			\midrule
			Top $1$ acc. $=\gamma_2^1$ (\%) &   $\mathbf{66.07}$ &  $56.06$ &  $61.40$ \\
			% \midrule
			% Top $2$ acc. $=\gamma_2^2$ (\%) &   $\mathbf{84.87}$ &  $76.97$ &  $81.31$ \\
			\midrule
			Top $3$ acc. $=\gamma_2^3$ (\%) &   $\mathbf{90.66}$ &  $85.09$ &  $88.77$ \\
			\toprule
			\textit{Sionna} &  RFF & MLP & $1$-NN\\
			\midrule
			Top $1$ acc. $=\gamma_2^1$ (\%) &   $66.07$ &  $54.07$ &  $\mathbf{69.73}$ \\
			% \midrule
			% Top $2$ acc. $=\gamma_2^2$ (\%) &   $75.13$ &  $65.00$ &  $\mathbf{79.47}$ \\
			\midrule
			Top $3$ acc. $=\gamma_2^3$ (\%) &   $78.27$ &  $69.07$ & $\mathbf{81.87}$ \\
			\bottomrule
		\end{tabular}
	\caption{Beam prediction accuracy}
	\label{table:res_acc_pseudolocs}
\end{table}
\noindent\textbf{Inference time.} It is proposed to compare the inference time of the different approaches in Table~\ref{table:inference_time}: $1$-NN ($1$) refers to the optimized ball-tree $1$-NN, while $1$-NN ($2$) refers to the brute force $1$-NN. One can see that the neural networks have faster inference times: this is due to their GPU implementation. Note that a GPU-optimized $1$-NN could be interesting as it has been shown in Table~\ref{table:res_acc_pseudolocs} that this method worked well with pseudo-locations. However, one has to keep in mind that, when considering online learning, the neural network approaches would outperform the $1$-NN in inference complexity, as for each new sample, the $1$-NN approach would have to compute an entire distance matrix whose size would grow with time.

\begin{table}[!h]
	\centering
	\begin{tabular}{lccccc}
		\toprule
		&  RFF & MLP & $1$-NN ($1$) & $1$-NN ($2$) \\
		\midrule
		Exec. time (ns) & $602.6$ &  $\mathbf{145.8}$ &  $4928.2$ &  $10913.9$ \\
		\bottomrule
	\end{tabular}
	\caption{Inference time}
	\label{table:inference_time}
\end{table}

\noindent\textbf{Precoder learning performance.} In this experiment, the goal is to learn the precoder from the pseudo-location. As opposed to the previous experiment, where the learning task was a classification task, this is a regression task. A network, called RFF (regr.) is defined as the same network than in Fig.~\ref{fig:RFF_net}, only dropping the last softmax non-linearity and adapting the last fully connected layer size so that the output is a precoding vector $\mathbf{w} \in \mathbb{C}^{N_a}$. The same holds true for the MLP baseline (MLP (regr.)). The same correlation-based training as in~\cite{LeMagoarou2022} is used. The $1$-NN baseline is defined as follows: for a given test pseudo-location, the associated precoder is the normalized downlink channel of the nearest training pseudo-location. The performance of the beam prediction networks (RFF (classif.) and MLP (classif.)) are also presented for this regression task: for a given pseudo-location each network infer a precoder in the codebook, and then its correlation with the downlink channel at the central frequency is computed (as in Eq.~\eqref{eq:corr}). The performance metric is the cumulative distribution function (CDF) of the normalized correlation between the learned precoder and the downlink channel at the central frequency.

\begin{figure}[t]
	\centering
	\includegraphics[width=.75\columnwidth]{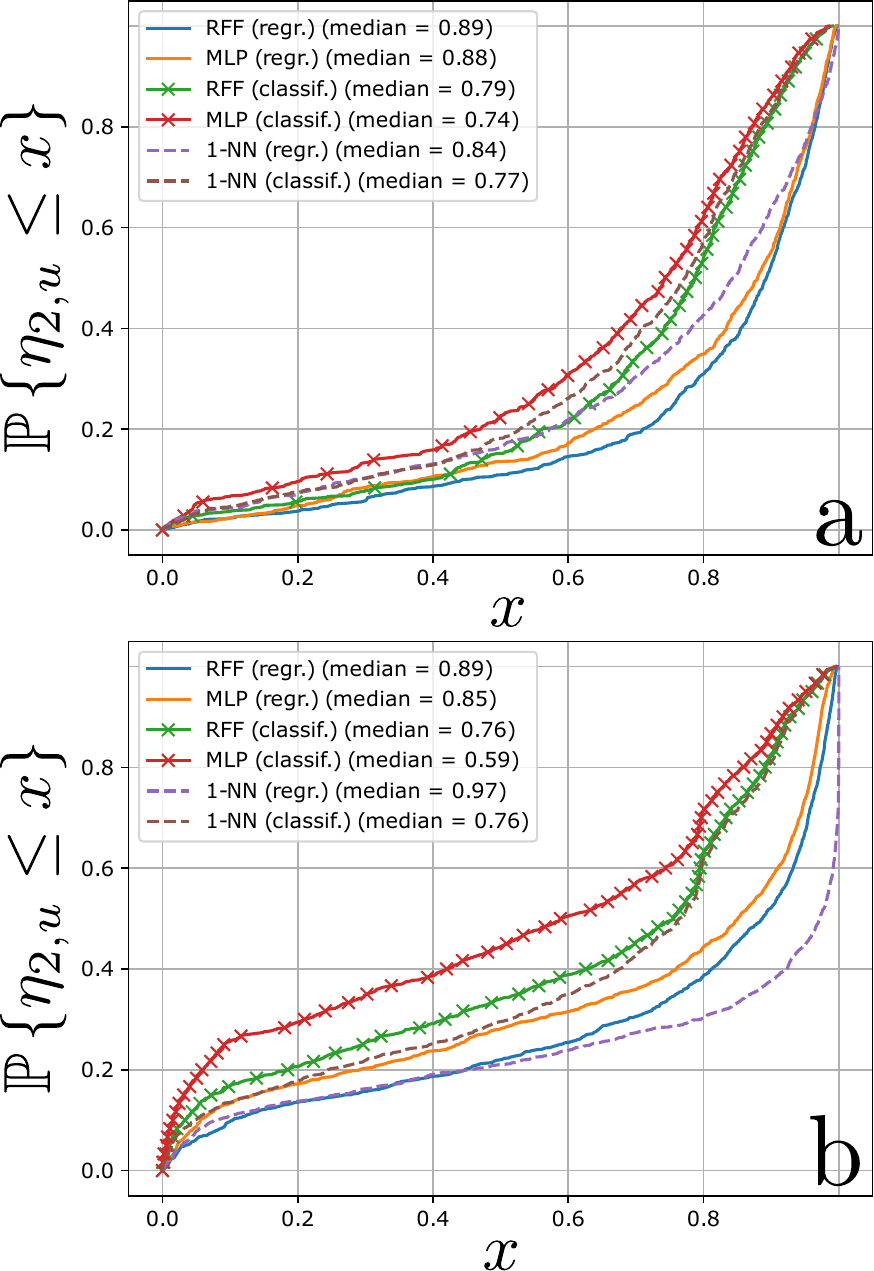}
	\caption{CDF of the correlations (BS$2$, a: \textit{DeepMIMO}, b: \textit{Sionna})}
	\label{fig:cdf}
\end{figure}

\begin{figure}[tb]
	\centering
	\includegraphics[width=.95\columnwidth]{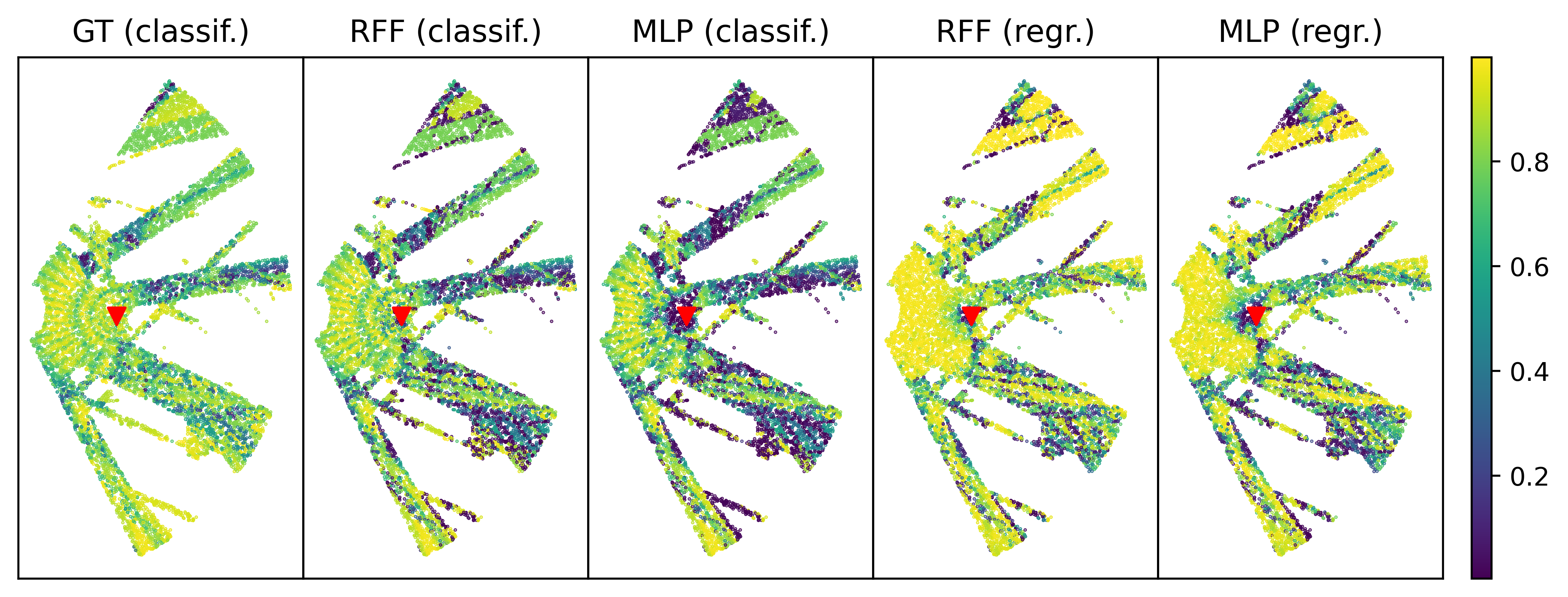}
	\caption{Correlation maps ($\eta_{2,u}$, BS$2$, \textit{Sionna})}
	\label{fig:correlation_map}
\end{figure}

One can observe in Fig.~\ref{fig:cdf} that, for both datasets, the regression networks (that learn the precoder) have better performances than the classification networks (that learn the best precoder in the codebook). This is easily explainable, as not having any constraints on the azimuth and elevation beam steering angle allows to learn the optimal precoder through regression networks, as opposed to the classification networks where angular constraints are defined by the codebook. Moreover, note that, in opposition to the codebook approach, the precoder learned through regression networks can be a linear combination of any steering vector. One can also note that, for the DeepMIMO dataset, RFF and MLP regression networks outperform the 1-NN approach. The performance gain of the regression networks can be seen in the normalized correlation maps of Fig.~\ref{fig:correlation_map}: regression networks offer a higher correlation coverage than classification networks.

% \begin{figure*}[!t]
%     \centering
%     \begin{subfigure}[t]{.4\textwidth}
% 		\centering
%         \includegraphics[width=\textwidth]{figs/DeepMIMO_cdf_comp_BS2.pdf}
%         \caption{DeepMIMO}\label{fig:subfig_cdf_deepmimo}
%     \end{subfigure}
%     \qquad
%     \begin{subfigure}[t]{.4\textwidth}
% 		\centering
%         \includegraphics[width=\textwidth]{figs/sionna_cdf_comp_BS2.pdf}
%         \caption{Sionna}\label{fig:subfig_cdf_sionna}
%     \end{subfigure}
% 	\caption{CDF of the normalized correlations (BS$2$)}
% 	\label{fig:cdf}
% \end{figure*}

\section{Conclusion and outlook}
\label{sec:conclusion}

In this paper, a model-based neural architecture was proposed to learn the mapping between a pseudo-location obtained through channel charting and the best beam in a codebook. Moreover, a performance comparison between networks that learn the best beam in a codebook (i.e. classification networks) and networks that directly learn a precoder (i.e. regression networks) was performed. The performance of the proposed architecture was evaluated on realistic synthetic data obtained from two different datasets. It has been shown that the proposed architecture outperforms the classical MLP architecture for the classification task. It has also been shown that, on those datasets, directly learning the precoder, rather than learning a precoder in a codebook, is beneficial in terms of performance. Finally, it has been shown that the low-dimensional pseudo-location is descriptive enough so that very simple methods such as the 1-NN obtain good performance. Future work will compare the performance of the proposed architecture when the pseudo-locations are obtained through the well-studied CSI-compression networks. Moreover, an end-to-end training is envisioned, where the neural network weights and biases and the matrices $\mathbf{D}$ and $\mathbf{Z}$ of Fig.~\ref{fig:inference_isomap} are jointly learned.

\bibliographystyle{./IEEEtran.bst}
\bibliography{./refs/paper_biblio.bib}

\end{document}